\documentclass[twocolumn,english,prl,showpacs]{revtex4}
\usepackage[T1]{fontenc}
\usepackage[latin9]{inputenc}
\usepackage{bm}
\usepackage{amsmath}
\usepackage{amssymb}
\usepackage{graphicx}

\makeatletter
\@ifundefined{textcolor}{}
{%
 \definecolor{BLACK}{gray}{0}
 \definecolor{WHITE}{gray}{1}
 \definecolor{RED}{rgb}{1,0,0}
 \definecolor{GREEN}{rgb}{0,1,0}
 \definecolor{BLUE}{rgb}{0,0,1}
 \definecolor{CYAN}{cmyk}{1,0,0,0}
 \definecolor{MAGENTA}{cmyk}{0,1,0,0}
 \definecolor{YELLOW}{cmyk}{0,0,1,0}
}

\makeatother

\usepackage{babel}
\begin{document}

\title{Strain manipulation of Majorana fermions in graphene armchair nanoribbons}

\author{Zhen-Hua Wang,$^{1,2}$ Eduardo V. Castro,$^{1,3}$ and Hai-Qing
Lin$^{1}$}

\affiliation{$^{1}$Beijing Computational Science Research Center, Beijing 100084,
China}

\affiliation{$^{2}$Institute for Quantum Science and Engineering and Department
of Physics, Southern University of Science and Technology, Shenzhen
518055, China }

\affiliation{$^{3}$CeFEMA, Instituto Superior T\'{e}cnico, Universidade de Lisboa,
Av. Rovisco Pais, 1049-001 Lisboa, Portugal}
\begin{abstract}
Graphene nanoribbons with armchair edges are studied for externally
enhanced, but realistic parameter values: enhanced Rashba spin-orbit
coupling due to proximity to a transition metal dichalcogenide like
WS$_{2}$, and enhanced Zeeman field due to exchange coupling with
a magnetic insulator like EuS under applied magnetic field. The presence
of s--wave superconductivity, induced either by proximity or by decoration
with alkali metal atoms like Ca or Li, leads to a topological superconducting
phase with Majorana end modes. The topological phase is highly sensitive
to the application of uniaxial strain, with a transition to the trivial
state above a critical strain well below $0.1\%$. This sensitivity
allows for real space manipulation of Majorana fermions by applying
non-uniform strain profiles. Similar manipulation is also possible
by applying inhomogeneous Zeeman field or chemical potential.
\end{abstract}


\maketitle


\emph{Introduction.---}Majorana fermions -- particles which are their
own anti-particles \cite{Wilczek09} -- have recently been the subject
of intense research due to the real prospect of realizing such exotic
particles in condensed matter platforms \cite{alicea12rev,beenakkerREv2013}.
Signatures of Majorana fermions have already been found in semiconducting
nanowires with strong spin-orbit coupling (SOC) in proximity to a
superconductor \cite{mourik2012signatures,deng2012anomalous,das2012zero,rokhinson2012fractional,PhysRevB.87.241401,finckPRL2013,dengSciRep2014,marcusNature2016},
at the end of atomic iron chains on the surface of a superconductor
\cite{nadj2014observation,meyer2016majorana}, and recently in the
hybrid system of a quantum anomalous Hall insulator coupled with a
superconductor as one-dimensional chiral modes \cite{majoranas2D2017}.
These \emph{condensed matter} Majorana fermions occur as zero energy
quasiparticles when topological superconductivity sets in the system.
These quasiparticles, so-called Majorana zero modes, obey non-Abelian
statistics \cite{ivanov2001,flensbergRev2012} and are seen as promising
building blocks to realize decoherence free topological quantum computation
\cite{RevModPhys.80.1083,dSFN15}. The underlaying condensed matter
support of the bounded Majorana zero modes will certainly play a key
role, dictating how easily Majorana fermions can be braided, or manipulated
in general. Graphene, with its highly tunable properties, is a tempting
platform. 

Apart from graphene's unconventional behavior \cite{RevModPhys.81.109}
and potential for applications in many different areas (energy \cite{bonaccorso2015graphene},
friction \cite{berman2015macroscale}, biology \cite{mohanty2008graphene},
to mention a few), it has not been recognized as an adequate host
for Majorana zero modes. A key ingredient in this context is the ability
to induce topological superconductivity in the host system \cite{QZrmp11,sau2010generic,aliceaPRB2010,PhysRevLett.105.077001,PhysRevLett.105.177002}.
This can be achieved by proximity to a normal superconductor and applied
magnetic field if the system has a sizable SOC, as is the case of
the semiconducting nanowires used in recent experiments \cite{mourik2012signatures,deng2012anomalous,das2012zero,rokhinson2012fractional,PhysRevB.87.241401,finckPRL2013,dengSciRep2014,marcusNature2016}.
Graphene has a vanishing SOC \cite{PhysRevB.74.155426,gmitra2009band},
so topological superconductivity seems to be ruled out based on the
standard ingredients. 

This state of affairs have changed completely in recent years, when
an enhancement of SOC by several orders was shown by two different
methods: proximity to high SOC transition metal dichalcogenides like
WS$_{2}$ \cite{NetoProximitySO2014,MorpurgoProximity2015,aliceaSO2015,morpurgoSOI2016},
and via graphene hydrogenation \cite{balakrishnan2013colossal,fabianSOh2013}.
Moreover, the other basic ingredient to achieve topological superconductivity
in the standard scheme, namely a magnetic field induced Zeeman field,
was recently shown to achieve values comparable to those in Rashba
nanowires like InAs, despite the much smaller $g$-factor in graphene.
This has been achieved through a strong magnetic exchange coupling
which occurs between graphene and the magnetic insulator EuS \cite{WLL+16},
clearly showing that graphene's tunability applies to a great variety
of parameters.

\begin{figure}
\begin{centering}
\includegraphics[width=1\columnwidth]{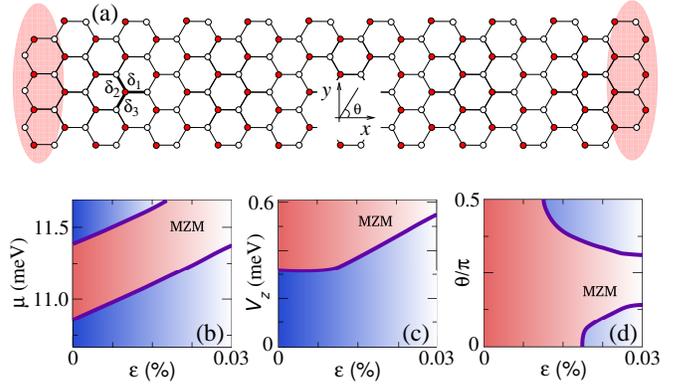}
\par\end{centering}

\caption{\label{fig:armchairPD}(a)~Sketch of a graphene armchair ribbon with
Majorana zero modes at both ends realized in the topological region
of the phase diagram~(b)-(d). Phase diagram in the plane of chemical
potential $\mu$ (b), Zeeman field $V_{z}$ (c), and strain direction
$\theta$ (d), vs strain $\varepsilon$. The topological phase with
Majorana zero modes corresponds to the MZM region in the figures.}
\end{figure}

In the present paper, we show that topological superconductivity can
be realized in armchair graphene nanoribbons {[}Fig.~\ref{fig:armchairPD}(a){]}
with externally enhanced, but realistic parameter values. As a two
dimensional membrane embedded in three-dimensional space, graphene
can easily develop built-in strain which we also consider here. Indeed,
built in strain of order $0.01\%-0.1\%$ has been reported in suspended
samples \cite{COK+10,OCK+12}. For supported samples, depending on
the substrate, this value can be even larger, as is the case of SiO$_{2}$
\cite{LLM+09}. Taking into account the effect of strain we obtained
the phase diagrams shown in Fig.~\ref{fig:armchairPD}(b)-\ref{fig:armchairPD}(d)
in the plane of uniaxial strain versus chemical potential $\mu$,
Zeeman field $V_{z}$, and strain direction $\theta$. The topological
phase is robust for realistic values of the parameters, and is also
very sensitive to strain, which can be used to tune the transition.
This leads to the possibility of real space manipulation of Majorana
zero modes by applying non-uniform strain, as is shown in the present
work.


\emph{Model and methods.---}We model $p_{z}$ electrons in graphene
with a tight binding Hamiltonian,
\begin{eqnarray}
H & = & -t\sum_{\mathbf{r},\bm{\delta},\sigma}c_{\mathbf{r},\sigma}^{\dagger}c_{\mathbf{r}+\bm{\delta},\sigma}+i\lambda\sum_{\mathbf{r},\bm{\delta}}\sum_{\sigma,\sigma'}c_{\mathbf{r},\sigma}^{\dagger}(\hat{\bm{\delta}}\times\bm{\sigma})_{z}^{\sigma\sigma'}c_{\mathbf{r}+\bm{\delta},\sigma'}\nonumber \\
 &  & +V_{Z}\sum_{\mathbf{r},\sigma,\sigma'}c_{\mathbf{r},\sigma}^{\dagger}\sigma_{\alpha}^{\sigma\sigma'}c_{\mathbf{r},\sigma'}+\Delta\sum_{\mathbf{r}}(c_{\mathbf{r},\uparrow}^{\dagger}c_{\mathbf{r},\downarrow}^{\dagger}+h.c.)\,,\label{eq:H}
\end{eqnarray}
where $c_{\mathbf{r},\sigma}^{\dagger}$ creates an electron with
spin $\sigma$ at site $\mathbf{r}$ of the honeycomb lattice, and
the three vectors $\bm{\delta}$ connect nearest neighbor atoms as
shown in Fig.~\ref{fig:armchairPD}(a). The first term in Eq.~\eqref{eq:H}
is graphene's minimum tight binding Hamiltonian with $t\approx3\,\mbox{eV}$
for the nearest neighbor hopping integral \cite{RevModPhys.81.109}.
The second term is the Rashba SOC \cite{KM05}, where we defined the
unit vector $\hat{\bm{\delta}}=\bm{\delta}/a$ with $a$ for the carbon-carbon
distance. The third term is the Zeeman coupling induced by an in-plane
magnetic field $\mathbf{B}=B\vec{e}_{\alpha}$ in the direction $\alpha=x,y$,
and the last term is the induced $s-$wave superconductivity, as in
Ref.~\cite{lossRotate2013}. 

Strain is introduced through changes on the matrix elements of the
Hamiltonian connecting nearest neighbor sites. For the hopping parameter
the change reads \cite{VKG10,friends},
\begin{equation}
t_{\hat{\bm{\delta}}}\rightarrow t_{\hat{\bm{\delta}}}(1-\beta(\hat{\bm{\delta}}\cdot\nabla)\mathbf{u}.\hat{\bm{\delta}}\,,\label{eq:hoppchange}
\end{equation}
where $\mathbf{u}$ is the deformation field, and $\beta=-\partial\ln t/\partial\ln a\approx2$.
Here we consider uniaxial strain applied along a direction specified
by $\theta$ {[}see Fig.~\ref{fig:armchairPD}(a){]}, and parametrize
the strain tensor $u_{ij}=(\partial_{j}u_{i}+\partial_{i}u_{j})/2$
appearing in Eq.~\eqref{eq:hoppchange} in terms of a single parameter
$\varepsilon$ \cite{PhysRevB.80.045401}. A similar strain induced
change can be used for Rashba SOC \cite{foot1}, but no quantitative
changes can be noticed in the phase diagram.

The topological and trivial phases are identified by: the presence
of a finite gap, with a gap closing point separating the two phases;
the presence or absence of zero energy states in finite length ribbons,
where their presence signals a topological state; the topological
invariant, which in this case can be chosen to be the Berry phase
$\gamma$ \cite{LCS16}, where $\gamma=\pi$ in the topological phase
and $\gamma=0$ in the trivial phase \cite{schnyder2008classification,RSFL10,hatsugaiBerry}.

\begin{figure}
\begin{centering}
\includegraphics[width=0.95\columnwidth]{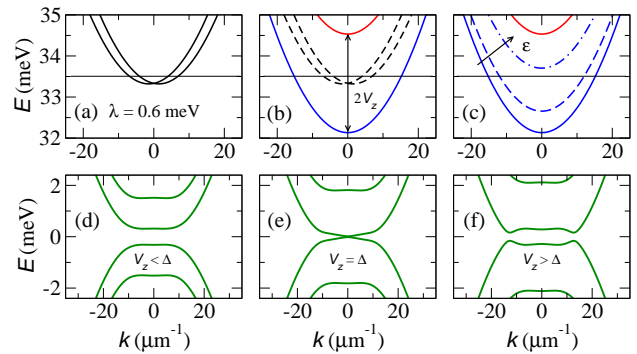}
\par\end{centering}

\caption{\label{fig:spectrumk}Lowest conduction band for an armchair ribbon
of width $N_{y}=81$ unit cells. The effect of Rashba SOC is shown
in (a), and the effect of Zeeman coupling in (b), while in (c) we
show the evolution of the lowest band with strain, $\varepsilon=0,\,0.01\%,\,0.03\%$.
For $\Delta\protect\neq0$ the system develops a gap, which closes
at $V_{z}=\Delta$ (d)-(f).}
\end{figure}


\emph{Topological superconducting phase.---}We consider the lowest
conduction band of an insulating armchair ribbon, as in \cite{lossRotate2013},
where the valley degeneracy is broken by boundary effects. As shown
in \cite{lossRotate2013}, and reproduced here for clarity, Rashba
SOC lifts spin degeneracy by displacing the parabolic bands horizontally
in opposite directions. The psectrum can be seen in Fig.~\ref{fig:spectrumk}(a)
for a conservative SOC value $\lambda\simeq0.6\,\mbox{meV}$ \cite{NetoProximitySO2014,MorpurgoProximity2015,aliceaSO2015,morpurgoSOI2016}.
The Zeeman coupling lifts the remaining spin degeneracy at $k=0$,
opening up a gap of value $2V_{z}$, as shown in Fig.~\ref{fig:spectrumk}(b).
Here we consider the realistic Zeeman field $V_{z}\simeq1.2\,\mbox{meV}$
(the experimental realization is discussed below). Strain has a strong
impact in the system, as shown in Fig.~\ref{fig:armchairPD}(b)-\ref{fig:armchairPD}(d),
and this is due primarily to strain induced changes in the band structure
of the ribbon. This is shown in Fig.~\ref{fig:spectrumk}(c) for
strains $\varepsilon=0,\,0.01\%,\,0.03\%$. In the present case strain
shifts the bands up, but depending on the ribbon width they can also
be shifted down \cite{foot2}. The effect on the topological phase
is the same.

\begin{figure}
\begin{centering}
\includegraphics[width=1\columnwidth]{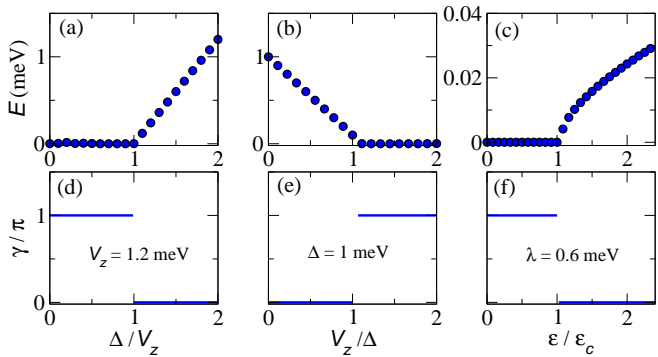}
\par\end{centering}

\caption{\label{fig:zeroEberry}Closest to zero energy mode for a finite ribbon
of length $N_{x}=10^{4}$ unit cells ($N_{y}=20$) as a function of
$\Delta$ (a), $V_{z}$ (b), and strain $\varepsilon$ (c). Berry
phase for the infinite ribbon as a function of $\Delta$ (d), $V_{z}$
(e), and strain $\varepsilon$ (f).}
\end{figure}

In the presence of a finite $s$-wave superconducting pairing $\Delta$
the system becomes gapped, as shown in Fig.~\ref{fig:spectrumk}(d)
for $V_{z}<\Delta$. The system goes through a gapless transition
point at $V_{z}=\Delta$ {[}Fig.~\ref{fig:spectrumk}(e){]} and becomes
gapped again for $V_{z}>\Delta$ {[}Fig.~\ref{fig:spectrumk}(f){]}.
In order to determine whether this is a topological transition we
have computed the low energy spectrum for a ribbon of finite length,
Fig.~\ref{fig:zeroEberry}(a)-\ref{fig:zeroEberry}(c), and the Berry
phase $\gamma$, Fig.~\ref{fig:zeroEberry}(d)-\ref{fig:zeroEberry}(f).
As can be seen in Figs.~\ref{fig:zeroEberry}(a) and \ref{fig:zeroEberry}(d),
or \ref{fig:zeroEberry}(b) and \ref{fig:zeroEberry}(e), the system
is topological as long as $V_{z}>\Delta$. For the realistic values
$V_{z}\simeq1.2\,\mbox{meV}$ and $\Delta\simeq1\,\mbox{meV}$ (experimental
realization is discussed below), it is clearly shown in Figs.~\ref{fig:zeroEberry}(c)
and \ref{fig:zeroEberry}(f) that a critical strain $\varepsilon_{c}$
exists above which the system becomes topologically trivial. For the
parameters used we obtain $\varepsilon_{c}\approx0.018\%$ for a ribbon
width of $\sim20\,\mbox{nm}$ ($N_{y}=81$), with $\varepsilon_{c}$
not very much dependent on the width ($\varepsilon_{c}\approx0.015\%$
for a ribbon with $N_{y}=20$). Such a small critical strain $\varepsilon_{c}$
indicates a high sensitivity to lattice deformations, which is primarily
due to the effect that strain has on the original band structure.
As shown in Fig.~\ref{fig:spectrumk}(b), if we start with the chemical
potential $\mu$ inside the $2V_{z}$ gap, as indicated by the horizontal
line, there is then a critical strain for which $\mu$ ceases to be
in the gap, as represented in Fig.~\ref{fig:spectrumk}(c). The dependence
of the critical strain $\varepsilon_{c}$ with chemical potential
$\mu$, Zeeman field $V_{z}$, and strain direction $\theta$, is
shown as the phase transition line in the phase diagrams of Figs.~\ref{fig:armchairPD}(b),
\ref{fig:armchairPD}(c), and \ref{fig:armchairPD}(d), respectively.
A high degree of tunability within realistic parameter values is apparent.

\begin{figure}
\begin{centering}
\includegraphics[width=0.98\columnwidth]{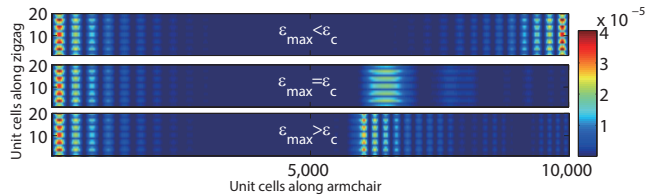}
\par\end{centering}

\caption{\label{fig:inhomoStrain}Effect of a strain profile in the Majorana
zero mode wave function for three different profile heights parametrized
by $\varepsilon_{max}$ according to Eq.~\eqref{eq:tanhProfile}.}
\end{figure}

\emph{Manipulating Majorana zero modes.---}The sensitivity to strain
found in this work can be used for spatial manipulation of Majorana
zero modes. This is demonstrated in Fig.~\ref{fig:inhomoStrain},
where the armchair ribbon has been subjected to a non-uniform strain
profile given by

\begin{equation}
\varepsilon=\varepsilon_{max}\frac{1}{2}[1+\tanh(\frac{x-x_{0}}{\zeta})]\,,\label{eq:tanhProfile}
\end{equation}
where $x$ is the coordinate along the armchair direction, as indicated
in Fig.~\ref{fig:armchairPD}(a). Each panel in Fig.~\ref{fig:inhomoStrain}
shows the Majorana zero mode wave-function amplitude along the ribbon
for different profile heights $\varepsilon_{max}$. If $\varepsilon_{max}<\varepsilon_{c}$,
the two Majorana zero modes appear localized at the two opposite ends
of the ribbon (top panel). At $\varepsilon_{max}=\varepsilon_{c}$
the left side of the ribbon ($x>x_{0}$) goes through a topological
phase transition, and the gap closes (middle panel). For $\varepsilon_{max}>\varepsilon_{c}$
the region $x>x_{0}$ becomes trivial, and one of the Majorana zero
modes localizes at $x_{0}$ (bottom panel). The Majorana zero mode
has thus been transferred to $x_{0}$ through strain manipulation.
We used $x_{0}=0.6aN_{x}$, with $N_{x}=10^{4}$, and $\zeta=0.5a$,
where $a$ is the lattice spacing in the armchair direction.

\begin{figure}
\begin{centering}
\includegraphics[width=0.98\columnwidth]{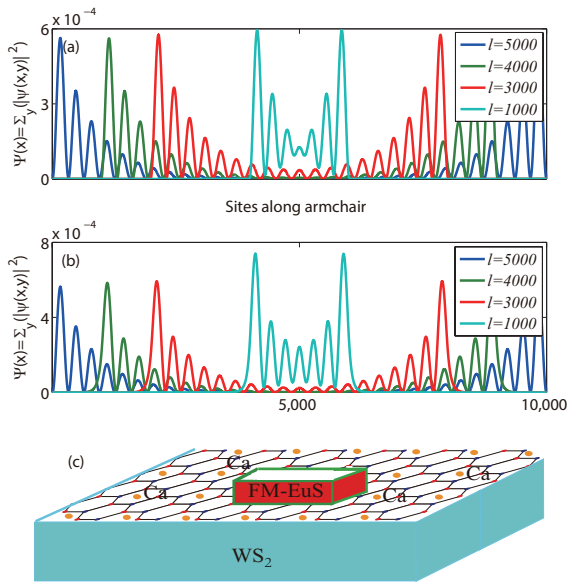}
\par\end{centering}

\caption{\label{fig:device}Effect of non-uniform $\mu$ (a) and $V_{z}$ (b)
in the Majorana zero mode wave function according to Eqs.~\eqref{eq:chemNonuniform}
and~\eqref{eq:VzNonuniform}, respectively. Parameters: $x_{0}=0.5aN_{x}$,
with $N_{x}=10^{4}$, and $\zeta=0.5a$, where $a$ is the lattice
spacing in the armchair direction. (c)~Device proposed in the present
work: graphene armchair nanoribbon sandwiched between high SOC transition
metal dichalcogenide WS$_{2}$ and a thin layer of ferromagnetic insulator
EuS, decorated with alkali metal atoms Ca or Li.}
\end{figure}

The advantage of graphene compared to other platforms is that not
only strain but also other parameters -- like the chemical potential
$\mu$ and the Zeeman coupling $V_{z}$ -- can be manipulated externally,
and non-uniform profiles can also be realized for these cases. In
particular, by applying a non-uniform chemical potential profile $\tilde{\mu}(x)$
according to 
\begin{equation}
\tilde{\mu}(x)=\bar{\mu}+\frac{\delta}{2}[\tanh(\frac{x-(x_{0}-l/2)}{\zeta})-\tanh(\frac{x-(x_{0}+l/2)}{\zeta})]\,,\label{eq:chemNonuniform}
\end{equation}
the topological region can be confined to a region of length $2l$
instead of the ribbon's full size. In Fig.~\ref{fig:device}(a) we
show the Majorana zero mode wave-function for different $l$ values.
As long as $l$ is large enough to avoid hybridization, it is clear
that the two Majorana zero modes localize at the ends of the region
of size $2l$. We used $\bar{\mu}=0.06\,\mbox{meV}$ and $\delta\simeq0.074\,\mbox{meV}$
in Eq.~\eqref{eq:chemNonuniform}, so that in the central region
the chemical potential $\mu=\bar{\mu}+\delta$ is inside the topological
gap. A similar effect is observed for a non-uniform Zeeman coupling
$\tilde{V}_{z}(x)$. This is shown in Fig.~\ref{fig:device}(b),
where a non-uniform profile given by 
\begin{equation}
\tilde{V}_{z}(x)=\frac{V_{z}}{2}[\tanh(\frac{x-(x_{0}-l/2)}{\zeta})-\tanh(\frac{x-(x_{0}+l/2)}{\zeta})]\,,\label{eq:VzNonuniform}
\end{equation}
was used.

\emph{Experimental realization.---}In order to realize Majorana zero
modes in graphene using the standard scheme, and then manipulate them
through strain, a sizable superconducting gap and Zeeman field are
required, and also needed is a high enough Rashba SOC. We show here
that such requirements are within experimental reach.

Regarding the necessary superconducting state, induced superconductivity
has been already demonstrated in graphene \cite{morpurgoSuperInduc2007}.
An interesting, tempting alternative is the recently discovered superconducting
state obtained by decorating graphene with the alkali metal atoms
Ca \cite{GeimSC} and Li \cite{ludbrookSupPNAS2015}. Superconductivity
has been shown to set in at $T_{c}\sim6\,\textrm{K}$, and the measurements
suggest a superconducting gap $\Delta\sim1\,\textrm{meV}$. Moreover,
in Ref.~\cite{klinovajaFloquet} it was shown that superconductivity
is even not needed if a driven electric field is applied.

Even though superconductivity and moderate to high magnetic fields
are usually not compatible, the 2D nature of graphene assures that
only the perpendicular component may weaken the superconducting state.
Such perpendicular component $B_{\perp}$ of the magnetic field $B$
can be made extremely small in graphene, with $B_{\perp}\lesssim50\,\textrm{mT}$
for $B=30\,\textrm{T}$, as shown in recent experiments \cite{geimLargeB}.
Furthermore, we can take advantage of graphene's tunability and use
the recently observed Zeeman field $V_{z}$ enhancement by one order
of magnitude \cite{WLL+16}, i.e. $V_{z}\sim1\,\textrm{meV}$ for
$B\gtrsim1\,\textrm{T}$, similar to nanowires with enhanced $g$-factor
like InAs \cite{marcusNature2016}, and fully compatible with the
superconducting gap mentioned above. This has been achieved by coupling
graphene to the magnetic insulator EuS \cite{WLL+16}. The effect
has been shown to be operative also on the surface of topological
insulators \cite{wei2013exchangeFITI} as well as in superconductors
\cite{hao1991thinEuS}. For thin films of EuS the system develops
an in-plane easy axis \cite{wei2013exchangeFITI}, and the effective
Zeeman coupling may then be tuned with an in-plane magnetic field,
which is the ideal configuration in the present case.

Finally, the Rashba SOC can be enormously enhanced by proximity to
WS$_{2}$ or other transition metal dichalcogenides, reaching values
$\lambda\sim1\,\textrm{meV}$ \cite{NetoProximitySO2014,MorpurgoProximity2015,aliceaSO2015}
or even higher \cite{morpurgoSOI2016}, in agreement with theoretical
calculations \cite{fabianSO2015,fabianSO2016}. Putting all pieces
together, we propose the device shown in Fig.~\ref{fig:device}(c)
to realize a topological superconducting state in graphene armchair
nanoribbons. Since WS$_{2}$ and graphene are two-dimensional materials,
and given that EuS can be grown in thin films, the effect of small
strain may be experimentally investigated using standard tools.

\emph{Conclusions.---}In the present work we have shown that topological
superconductivity can be realized in armchair graphene nanoribbons
with externally enhanced parameter values -- but still within realistic
conditions -- using the standard scheme with Rashba SOC, Zeeman field,
and $s$-wave superconductivity. A very high sensitivity to strain
has been put forward. A non-uniform strain profile can be used to
manipulate the position of the localized zero energy Majorana modes.
A spatial variation of the chemical potential or Zeeman field can
also be used to manipulate Majorana zero modes. Such non-uniform variation
of parameters can in principle be easier to achieve in graphene than
in other platforms. 

It would be interesting to investigate the impact of strain in two
recent proposals for realizing Majorana fermions in graphene where
SOC enhancement is not required \cite{lossRotate2013,sanJosePRX2015}.
In Ref.~\cite{lossRotate2013}, a spatially varying magnetic field
was shown to give rise to an additional term in the Hamiltonian which
is equivalent to SOC, thus mitigating the lack of SOC in graphene.
In Ref.~\cite{sanJosePRX2015}, the interaction-induced magnetic
ordering of graphene's zero Landau level is shown to give rise to
topological superconductivity when the graphene edge is in proximity
to a conventional superconductor. No SOC is required in that case.
The impact of strain in these two setups may provide an interesting
way to manipulate Majorana zero modes.
\begin{acknowledgments}
Z.-H.W. acknowledges support from China Postdoctoral Science Foundation
(2016M591057). E.V.C. acknowledges partial support from FCT-Portugal
through Grant No. UID/CTM/04540/2013. H.Q.L. and Z.-H.W. acknowledge
financial support from NSAF U1530401 and computational resources from
the Beijing Computational Science Research Center.

\end{acknowledgments}

\bibliographystyle{apsrev}

\begin{thebibliography}{60}
\bibitem[{foo({\natexlab{a}})}]{foot1}
\bibinfo{note}{We have used $\beta'=-\partial\ln\lambda/\partial\ln a\approx2$,
  but this value is largely unknown.}
\bibitem[{foo({\natexlab{b}})}]{foot2}
\bibinfo{note}{Uniform strain shifts the energy at the M points differently.
  For tensile strain along armchair, and considering the conduction band, two M
  points become lower energy while the other increases its energy. In the
  ribbon, the projected bands can go up or down in energy depending on the type
  of M points they are closer to.}
\end{thebibliography}

\end{document}